\pdfoutput=1

\documentclass[%
reprint,
groupedaddress,
 amsmath,amssymb,
aps,
prl,
floatfix,
]{revtex4-1}

\usepackage{graphicx}
\usepackage{hyperref}

\usepackage{natbib}
\begin{document}

\preprint{APS/123-QED}

\title{Multilayer microwave integrated quantum circuits for scalable quantum computing}


\author{T.~Brecht}
\email{teresa.brecht@yale.edu}
\author{W.~Pfaff} 
\author{C.~Wang}
\author{Y.~Chu}
\author{L.~Frunzio}
\author{M.H.~Devoret}
\author{R.J.~Schoelkopf}
\affiliation{Department of Applied Physics, Yale University, New Haven, Connecticut 06511, USA}

\date{\today}

\begin{abstract}
	As experimental quantum information processing (QIP) rapidly advances, an emerging challenge is to design a scalable architecture that combines various quantum elements into a complex device without compromising their performance.  In particular, superconducting quantum circuits have successfully demonstrated many of the requirements for quantum computing, including coherence levels that approach the thresholds for scaling.  However, it remains challenging to couple a large number of circuit components through controllable channels while suppressing any other interactions.  We propose a hardware platform intended to address these challenges, which combines the advantages of integrated circuit fabrication and long coherence times achievable in three-dimensional circuit quantum electrodynamics (3D cQED).  This multilayer microwave integrated quantum circuit (MMIQC) platform provides a path toward the realization of increasingly complex superconducting devices in pursuit of a scalable quantum computer.
\end{abstract}

\maketitle

Experimental quantum information processing is rapidly developing in several physical implementations, and superconducting quantum circuits are a particularly promising candidate for building a practical quantum computer.\cite{Devoret:2013jz,Barends:2014fu}  In these systems, qubits made with Josephson junctions behave like macroscopic atoms with quantized energy levels in the microwave domain.  Coupling them to resonators forms a powerful platform known as circuit quantum electrodynamics (cQED)\cite{Blais:2007hh,Schoelkopf:2008cs} that shares several important advantages with classical computing architectures:  For one, devices are created in the solid state and their properties can be fully engineered through circuit design and mass produced by lithographic fabrication.  Further, electromagnetically coupling qubits to superconducting transmission lines enables communication of quantum information, rapid multi-qubit gates, and entangling operations between elements on or off the chip.\cite{Barends:2014fu,Reed:2012hu,Siddiqientanglement, MajerNature2007,Steffen:2013ch,vanLooScience2013} Finally, electronic control and measurement are achieved through microwave signals carried to and from the device by wires and cables. This capacity for quantum electrical engineering of devices allows the lifetimes of quantum states to continually rise due to improvements in design and selection of high quality materials. As a consequence, superconducting circuits fulfill many of the necessary requirements for universal quantum computation, as evidenced by recent experimental realizations of a large suite of desired building blocks.\cite{Divincenzo2,Ladd:2010kq,PerezDelgado:2011jb}

Building a fully functional, fault-tolerant quantum computer, however, will require scaling to a device with orders of magnitude more circuit elements than today's experimental devices. When scaling up the system size, error rates need to remain below the threshold for quantum error correction.\cite{Gottesman:2009ug,Anonymous:NJHxykQQ} At the same time, the different components must retain the ability to selectively interact with each other while being externally addressed and accurately controlled. Finally, they must be mass producible in a reliable and precise manner. These criteria cannot simply be achieved by replicating and connecting currently available hardware.

Superconducting quantum circuits are presented with crucial challenges that prevent a scaling strategy similar to that of classical integrated circuits. The strong electromagnetic interactions of the qubits allow for efficient entanglement and control, but make them also especially susceptible to undesired couplings that degrade quantum information. This ``crosstalk" results in either undesirable mixing of quantum states or decoherence. High isolation to prevent these effects is especially important as high-Q qubits ($ \mathrm{Q} \approx 10^6 - 10^9$) must also be coupled to high-speed, low-Q ($\mathrm{Q} \approx 10^3$) elements for readout, control, and feedback. Various strategies are employed in large scale detector arrays and complex cQED circuits to mitigate crosstalk.\cite{ Noroozian:2012iy, Bintley:2012is, Zmuidzinas:2012kh,  Wenner:2011hf, Chen:2014he, Vesterinen:2014wk} On a small scale, these unintended couplings can be minimized by spectral and/or spatial separation among elements on a single chip. However, with increasing number of elements, the former approach faces an increasingly crowded spectrum. The latter becomes ineffective when the device package grows in size and consequently supports unintended electromagnetic modes that can mediate detrimental couplings between elements. Simply increasing the density of elements through miniaturization is subject to a trade-off between size and coherence due to dissipative materials surfaces.\cite{Moore1965, Wenner:2011co,Geerlings:2012ps,Reagor:2013tq,Wang:2015}

Scaling up quantum circuits also encounters a challenge in connectivity. A large number of qubits and resonators have to be selectively coupled to each other with very low loss.  This requires more internal connections between circuit elements than can be achieved in a single plane, demanding signal crossovers that are generally hard to implement with high isolation.\cite{Chen:2014he, JeffreyPRL2014}  Even in planar architectures that only require nearest-neighbor coupling, such as the surface code\cite{FowlerPRA2012}, additional connections are needed for measurement and control purposes. As a result, development towards low-loss three-dimensional connections has emerged as a pressing need. A superconducting quantum computer will also require a scalable input/output interface that provides external connections with much higher density and isolation than techniques that are presently employed, such as wire bonding.

Therefore, the outstanding task is to design a hardware platform that allows large number of quantum components to couple through intended and controllable channels, while suppressing any other interactions.  In addition, crucial properties of the quantum circuit, such as qubit and resonator frequencies, anharmonicities, and mutual interaction strengths must be more predictable and reproducible than what has been necessary to date.

It is possible to construct a scalable quantum computer that overcomes these challenges by incorporating other technologies into the cQED architecture. This can be achieved by dividing the quantum circuit into subsystems and shielding each one with a three-dimensional superconducting enclosure, as sketched in Figure 1. The enclosures, represented as boxes in the figure, suppress unintended crosstalk by providing a high degree of electromagnetic isolation. Some contain input and output circuitry, while others are 3D cavity resonators coupled to qubits.\cite{Brecht:APL} In this 3D geometry, both qubits and resonators have been shown to experience substantial gains in coherence times.\cite{Paik:2011hd} The use of intentionally designed 3D modes as quantum elements also alleviates the need for suppressing spurious package modes. Within each shielded module, circuit complexity can be kept manageable, and the engineering of different modules can be mostly independent. Interconnects between individual modules are then created using superconducting transmission lines. 3D shielding allows for the implementation of a large number of internal and external connections, shown as wires in Figure 1, without introducing crosstalk or sacrificing coherence. A dilution refrigerator payload space of one cubic meter can accommodate a device containing millions of these components, each with linear dimensions on the order of one centimeter.

\begin{figure}[hb]
\includegraphics[scale = 1,angle=0]{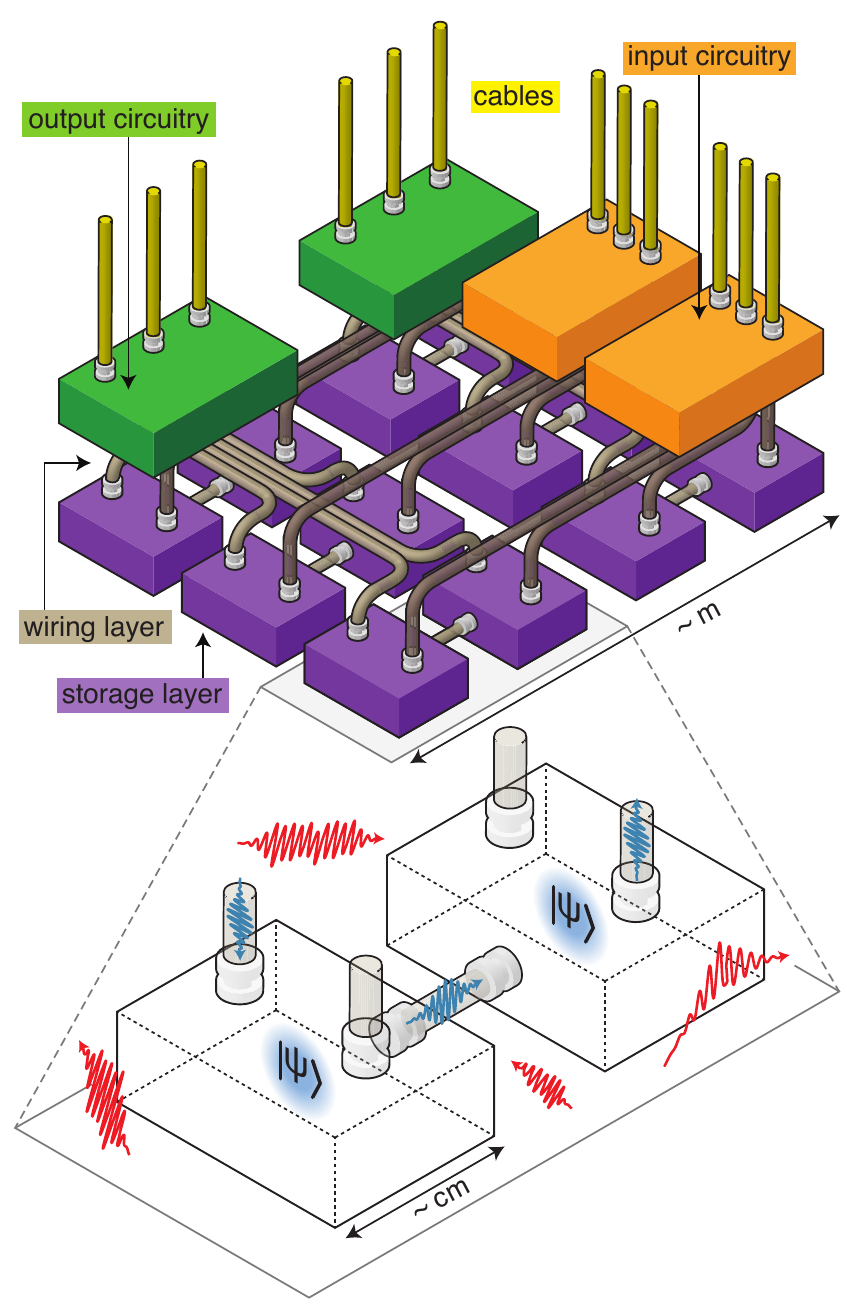}
\caption{
	\textbf{Conceptual sketch of a QIP device suitable for scaling.}  
	Modules of cQED components are surrounded by superconducting enclosures approximately a centimeter in size (boxes). Individual quantum states reside in storage modules with minimal loss and high isolation. Modules of input circuitry may include filtered RF control lines and bias wiring, and output circuitry may include quantum limited amplifiers, filters and switches. Key requirements of the device are addressability via a large number of external connections, indicated by yellow tubes at the surface, and internal selective coupling and isolation, indicated by grey tubes. Each component is well isolated from cross-talk and environmental effects (red wave packets) by its enclosure, and components interact only by transmission through shielded channels (blue wave packets). While this sketch shows only a few modules, a quantum processor composed of millions of these components can occupy the payload space of industry standard dilution refrigerators.
	}
\label{Fig1}
\end{figure}

As an approach for implementing this concept, we introduce the multilayer microwave integrated quantum circuit (MMIQC), which combines the advantages of integrated circuit fabrication with the long coherence times attainable in 3D cQED  (Figure 2). The shielding enclosures are formed using established techniques from the MEMS industry.\cite{MEMSbook} For example, recesses can be created in silicon wafers by masking and subsequent etching. The removal of bulk substrate material to create 3D features with precise dimensions is known as ``micromachining''. The etched wafers are then lithographically patterned with metal, aligned, and bonded to one another to form integrated circuitry and shielding.

\begin{figure*}[ht]
\includegraphics[scale = 1,angle=0]{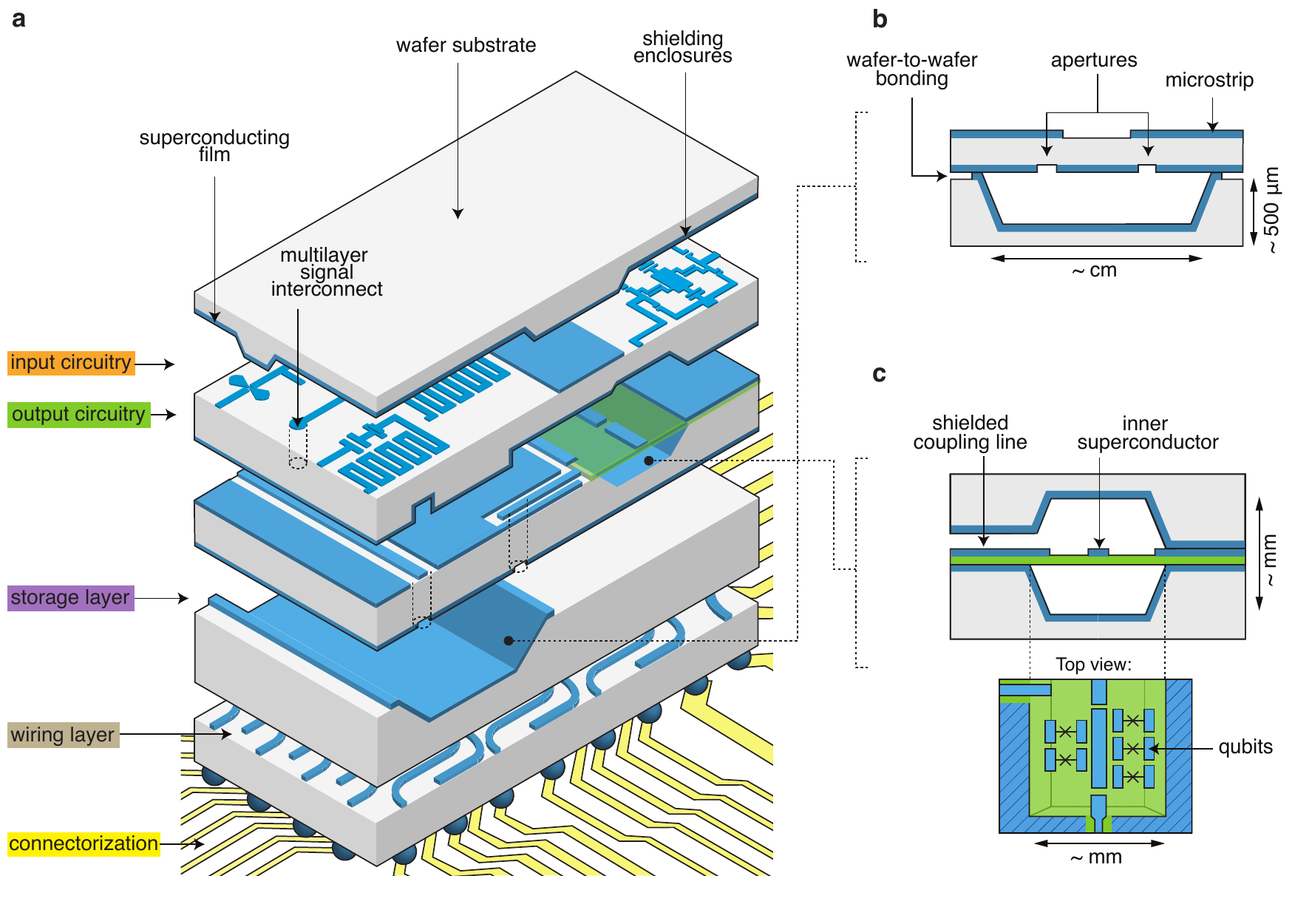}
\caption{
	\textbf{Proposed schematic of a multilayer microwave integrated quantum circuit (MMIQC).}  
	\textbf{a,} Layers are made of silicon wafers with features etched out of the bulk using micromachining techniques to create enclosures that serve as high-Q resonators and as shielding for internal components. Superconducting metalization (blue) covers the walls of these enclosures and enables low-loss wafer-to-wafer bonding of the layers.  Dense planar input/output circuitry, including filters, switches and amplifiers can be embedded in other layers.  RF and DC vertical interconnects carry signals between layers, and wirebonding or ball-grid connection can be used to interface with external control and measurement electronics. 
	\textbf{b,} A cross-section of the rectangular cavity resonator shows interlayer aperture coupling between the cavity and transmission lines above.  
	\textbf{c,} 3D superconducting transmission lines could be constructed using membranes (green) in the micromachined structure. Some of these can be populated with qubits and act as a compact low-loss quantum bus.
	}
\label{Fig2}
\end{figure*}

The MMIQC hardware platform is suited for integrating a wide variety of cQED components and utilizes a number of existing technologies.\cite{KatehiLPB:2001dl} First, high-Q resonators for quantum memories with precise frequencies are implemented with some of the superconducting enclosures (Figure 2b). Micromachined cavity resonators with Q$\sim$1000 have previously been demonstrated with normal metal coatings, and are used in low-loss multi-cavity microwave filters.\cite{Harle:2003tc,Brown:1998uz,Papapolymerou:1997fo} These existing methods can be extended to accommodate superconducting coatings to achieve the higher Q required for storing quantum states.\cite{Brecht:APL} 

Another important feature of the MMIQC is the use of shielded, low-loss, superconducting transmission lines connecting elements on the same layer (Figure 2c) and vertical interconnects to provide coupling between layers. Micromachining has been used to remove substrate in both normal metal and superconducting transmission lines to reduce dielectric loss.\cite{Brown:1998uz, Bruno:2015up} Several styles of low-loss micromachined transmission lines have been developed in normal metals and feature 3D shielding that can be adapted using superconducting materials.\cite{Katehi:1997wc, LlamasGarro:2003wl, Reid:2006us} 
To route signal communication between layers, MMIQCs must contain superconducting versions of layer-to-layer RF interconnects.\cite{Farrington:2011fm,Herrick} Coupling of the cavity resonators to one another or to planar transmission lines is achieved, for instance, through apertures in the metalization of the enclosure through which electromagnetic field radiates, as implemented in microwave cavities and multi-cavity filters.\cite{Harle:2003tc, Brown:1998uz, Papapolymerou:1997fo} In the MMIQC, superconducting transmission lines and interconnects based on these existing technologies will be further engineered to provide the large range of necessary couplings and minimize any parasitic losses.

Other existing quantum circuit elements of various geometries can be readily incorporated into the MMIQC. These include planar devices such as qubits and Josephson-junction based quantum amplifiers, which can be situated inside shielding enclosures on non-metalized surfaces. As shown in Figure 2c, some modules or buses coupled to several qubits can even be made on membranes to further suppress dielectric loss, as was demonstrated in normal metal filters.\cite{Chi:1996io,Blondy:1998uz} It is also possible to incorporate devices such as superconducting whispering gallery mode resonators.\cite{Minev:2013fd, Minev:2015} The device described in Ref. \cite{Minev:2015} demonstrates coupling of a planar fabricated qubit to the fields of a multilayer transmission line resonator, and the same mechanism can be applied to mediate coupling of fields between a variety of planar and 3D structures throughout the MMIQC.

The architecture presented here has several other advantages for meeting demands that future, large-scale QIP devices will face. First, it allows for high-density connectivity to external measurement and control circuitry through methods commonly used in integrated circuits, such as ball grid arrays and flip-chip bonding.\cite{Miller:2012vw, FlipChipBook}  Moreover, the platform is compatible with integration of classical control electronics in close proximity to the quantum device, for example to reduce latency in feedback circuits.  When incorporated with adequate isolation, such ``cold electronics'' may even be included in the same structure as the quantum elements.  Lastly, devices like the one shown schematically in Figure 2 can be fabricated in a single foundry to produce a compact monolithic structure of macroscopic size containing components made with lithographic precision.

While many features and components of the MMIQC make use of existing technologies, there are certain aspects of the design that will require the development of novel techniques. These include the fabrication of micromachined superconducting enclosures (both for use as cavity resonators and as shielding), and engineering coupling between 3D modes and embedded planar circuitry throughout the multilayer device. In particular, enclosures with high quality RF isolation will require a low-loss bond between two wafers that are etched and coated with a suitable superconductor. We have begun to investigate the feasibility of such devices by fabricating and testing 3D superconducting cavity resonators. As described in Ref. \cite{Brecht:APL}, we find that many of the above challenges can be addressed through the selection of appropriate methods and materials. 

In summary, we believe that practical implementation of complex quantum circuits will require innovative approaches to scaling up.  In the platform of cQED, the goal is to create integrated QIP circuits with multiple interconnected cavities, qubits, and embedded control wiring while also obeying the principles for minimizing materials losses, and providing the high isolation and shielding that quantum states require.  We emphasize that the most pressing need in achieving this goal is not the miniaturization of circuits, but the maintenance of high coherence.  We conclude that the solution is likely to heavily employ superconducting 3D enclosures.

We have proposed a lithographic approach for fabricating complex quantum circuits that incorporate superconducting enclosures. In a proof-of-principle experiment described in Ref. \cite{Brecht:APL}, we have demonstrated a 3D micromachined resonator with a high quality superconducting seam and planar multilayer coupling, which are two key components of the architecture presented here. We foresee that various other components including qubits and amplifiers can be readily incorporated in this multilayer geometry. Transmission lines, wiring, and filtering can also be added to provide suitably shielded connections between subsystems and to the outside world.  The eventual complete demonstration of the MMIQCs described here could enable the large-scale fabrication of sophisticated systems for quantum information processing.

We thank Harvey Moseley, Zlatko Minev, Ioan Pop, Kyle Serniak, and Michael Hatridge for useful conversations.  This research was supported by the U.S. Army Research Office (W911NF-14-1-0011). All statements of fact, opinion or conclusions contained herein are those of the authors and should not be construed as representing the official views or policies of the U.S. Government. W.P. was supported by NSF grant PHY1309996 and by a fellowship instituted with a Max Planck Research Award from the Alexander von Humboldt Foundation. 




\end{document}